\documentclass[12pt]{article}
\usepackage{amsthm,amsmath,bm,dsfont,amssymb,graphicx,multirow}
\usepackage{multicol}        
\usepackage{enumerate,verbatim}
\usepackage{natbib}
\newcommand{\indicator}[1]{\mathds{1}_{\left[ {#1} \right] }}
\newcommand{\argmin}{\operatornamewithlimits{argmin}}
\usepackage{url} 

\newcommand{\blind}{0}

\begin{document}

\def\spacingset#1{\renewcommand{\baselinestretch}%
{#1}\small\normalsize} \spacingset{1}


\if0\blind
{
  \title{\bf Mixtures of multivariate generalized linear models with overlapping clusters}
  \author{Saverio Ranciati\thanks{Corresponding author: Saverio Ranciati, Department of Statistical Sciences, University of Bologna, e-mail: saverio.ranciati2@unibo.it}\\
    Department of Statistical Sciences, University of Bologna\\
    and \\
    Veronica Vinciotti\\
    Department of Mathematics, Brunel University London\\
    and \\
    Ernst C. Wit\\
    Institute of Computational Science, Universit\`a della Svizzera italiana\\
    and \\
    Giuliano Galimberti\\
    Department of Statistical Sciences, University of Bologna}
    \date{}
  \maketitle
} \fi

\if1\blind
{
  \bigskip
  \bigskip
  \bigskip
  \begin{center}
    {\LARGE\bf Mixtures of multivariate generalized linear models with overlapping clusters}
\end{center}
  \medskip
} \fi

\bigskip
\begin{abstract}
With the advent of ubiquitous monitoring and measurement protocols, studies have started to focus more and more on complex, multivariate and heterogeneous datasets. In such studies, multivariate response variables are drawn from a heterogeneous population often in the presence of additional covariate information. In order to deal with this intrinsic heterogeneity, regression analyses have to be clustered for different groups of units.
Up until now, mixture model approaches assigned units to distinct and non-overlapping groups. However, not rarely these units exhibit more complex organization and clustering. It is our aim to define a mixture of generalized linear models with overlapping clusters of units. This involves crucially an overlap function, that maps the coefficients of the parent clusters into the the coefficient of the multiple allocation units.
We present a computationally efficient MCMC scheme that samples the posterior distribution of the parameters in the model. An example on a two-mode network study shows details of the implementation in the case of a multivariate probit regression setting. A simulation study shows the overall performance of the method, whereas an illustration of the voting behaviour on the US supreme court shows how the 9 justices split in two overlapping sets of justices.
\end{abstract}

\noindent%
{\it Keywords:} Multivariate generalized linear models; heterogeneity; mixture models; overlapping clusters; Bayesian inference; two-mode networks; probit regression.
\vfill

\newpage
\spacingset{1.5} 

\section{Introduction}
\label{sec:intro}

Clustering approaches are popular in many fields as they allow to identify unknown grouping structures from multivariate data. In a regression context, mixtures of Generalized Linear Models (GLMs) provide a natural model-based approach to account for the heterogeneity in the data due to the presence of heterogeneous clusters. These models have been developed extensively in the case of a single response variable, i.e., via a mixture of univariate GLMs \citep{grun2008finite}. However, with the advent of complex multivariate data both at the level of predictors and responses, multivariate regression models are now common in many fields \citep{fahrmeir2013multivariate}. Mixtures of multivariate GLMs were introduced by \cite{wedel1995mixture} with an implementation provided in the \texttt{R} package \texttt{FlexMix} \citep{JSSv028i04}. Further extensions to the high dimensional case have been studied recently by \cite{price2017cluster} using penalised inferential approaches.


In the traditional formulation of a mixture model, a unit can belong to only one of the clusters --- and therefore be modeled by only one component --- thus limiting the way we can build and discover more complex grouping structures of the units. In some applications, one can imagine that a statistical unit can be assigned simultaneously  to more than one cluster (\emph{multiple allocation}) or perhaps even to none of them. Some extensions to allow for overlapping clusters have been proposed with respect to conventional mixtures of exponential family distributions and mixtures of univariate regression models. For example, in \cite{banerjee2005model}  and  \cite{fu2008multiplicative} the model-based clustering strategy is modified to allow for components in the mixture that arise as a product of densities from the exponential family, inducing a resulting density that is still within the same exponential family. \cite{heller2007nonparametric} extended this approach in a nonparametric Bayesian fashion to allow for the selection of the number of components of the mixture. In both cases, however, the way the multiple allocation is handled is reflected into a strict definition of the parameters of the resulting cluster, which are limited by the mathematical derivation of the product of the densities. This hinders the flexibility of the model and the interpretability of the regression coefficients and related parameters. An alternative view to account for overlap is that of re-parameterizing the model in such a way that parameters of the overlapping clusters are linked to those of the originating clusters. This idea was explored by \cite{Ranciati2017chipseq} in the context of univariate mixture models and by \cite{ranciati2017identifying} in the context of network data.

In this paper, we take this alternative view forward and  propose a novel definition of and inferential strategy for a mixture of multivariate GLMs with overlapping clusters of units. In detail, Section~\ref{sec:multiGLMs} introduces multivariate GLMs and the definition of overlapping mixture components. We discuss the Bayesian implementation of our proposal, including cluster allocation and model selection. In Section~\ref{sec:probit} our general proposal is worked out for the case of multivariate binary observations, motivated by actor-event network data where explanatory variables may be available at the level of actors and/or events, and where overlapping clusters of actors may occur naturally. Section~\ref{sec:simul} shows the computational and inferential performance of our method through a simulation study, whereas Section~\ref{sec:supreme} illustrates the method on a two-mode network related to rulings from the Supreme Court of the United States of America.


\section{Mixture of multivariate GLMs with overlap}
\label{sec:multiGLMs}
In this section, we describe our proposal for a mixture of multivariate GLMs that can accommodate overlapping clusters. We call this \texttt{miro}, \textbf{mi}xture of \textbf{r}egression models with \textbf{o}verlap. We start by defining a mixture of multivariate generalized linear models. Suppose that we have $n$ multivariate observations measured on a $d$-dimensional space, so that each response vector $\mathbf{y}_i=(y_{1i}, \dots, y_{id})$ corresponds to a row of the $n\times d$ data matrix $y$. 
The aim is to cluster these $n$ observations based on their $d$ features, while accounting for possible additional information, either at the level of the units $i=1,\ldots,n$ or of the variables $j=1,\ldots,d$.
In a mixture of regression framework, the assumption is that the response variables come from a mixture of $K$-components (also called clusters) from the same multivariate exponential family, i.e.,
\[ \mathbf{y}_{i} \sim \sum_{k=1}^K \alpha_k \text{MVExpFam}\bigl(\boldsymbol{\theta}_{k}(\boldsymbol{x}_i, \boldsymbol{W});  \boldsymbol{\Sigma}_k \bigl),\nonumber \]
where $\boldsymbol{\alpha}=(\alpha_1,\dots, \alpha_K)$ are the prior cluster probabilities,  $\boldsymbol{\theta}_{k}(\boldsymbol{x}_i, \boldsymbol{W})$ is the natural parameter of the GLM and a function of cluster-specific parameters and covariate information, which could be unit-specific, $\boldsymbol{x}_i=(x_{i1},\ldots,x_{iL})$, or response variable-specific, $\boldsymbol{w}_j=(w_{j1},\ldots,w_{jd})$. Furthermore, depending on the distribution, $ \boldsymbol{\Sigma}_k$ refers to additional nuisance parameters, such as the dispersion parameters. An alternative hierarchical representation of the mixture model is achieved by introducing a unit-specific binary latent vector $\mathbf{z}_i=(z_{i1},\dots,z_{iK})$, made up of all zeros with the exception of a single element $z_{ik}=1$ for unit $i$ belonging to cluster $k$. 
In particular, using this notation, the hierarchical formulation of a mixture of multivariate GLMs is given by
\begin{eqnarray}\label{hier1}
\mathbf{z}_i &\sim& \text{Multinomial}(\alpha_1,\dots,\alpha_K) \\
\nonumber \mathbf{y}_i | \mathbf{z}_i, \boldsymbol{\theta}, \boldsymbol{\Sigma} &\sim& \prod_{k=1}^K \bigl[ \text{MVExpFam}\bigl(\boldsymbol{\theta}_{k}(\boldsymbol{x}_i, \mathbf{W});  \boldsymbol{\Sigma}_k \bigl)\bigl]^{z_{ik}}.
\end{eqnarray}

For example, if the response variables are binary outcomes denoting the attendances of $n$ units to $d$ events, the $n\times L$ matrix $ \boldsymbol{X}$, with i-th row $\boldsymbol{x}_i$,  collects $L$ characteristics of the $n$ units, such as age, gender and education, whereas the $d\times Q$ matrix $ \boldsymbol{W}$ describes the $Q$ features of the $d$ events, such as time, location and type of event. This is indeed the specific case that will be discussed more in detail in Sections~\ref{sec:probit} and \ref{sec:supreme}.

In finite mixtures of GLMs \citep{fahrmeir2013multivariate}, the parameters of each component are linked to the covariates via an appropriate link function applied to a linear combination of the predictors. Here we consider both individual covariates $\mathbf{x}_i$ and response-specific covariates $\mathbf{w}_j$. In particular, the natural parameter for individual $i$ part of cluster $k$ is given by $\theta_{kij} \equiv\theta_{k}(\mathbf{x}_i,\mathbf{w}_j) =g(\eta_{kij})$ with $g(\cdot)$ a link function and $\eta_{kij}$ the linear predictor. In a multivariate general setting, where covariates are available both at the level of units and responses, we specify each component of the $d$-dimensional vector $\boldsymbol{\eta}_{i}$ by
\begin{equation}\eta_{kij}=\mu_{k}  +  \mathbf{x}_i\boldsymbol{\beta}_{k}^{\intercal}   +  \boldsymbol{w}_j\boldsymbol{\gamma}_{k}^{\intercal}
\label{lin_pred1}
\end{equation}
where $\{\mu_k, \boldsymbol{\beta}_k, \boldsymbol{\gamma}_k  \}$ are cluster-specific vectors of parameters. In particular, for each cluster $k$, the model includes an intercept $\mu_k$, an $L$-dimensional vector of regression coefficients $\boldsymbol{\beta}_k$ for the unit covariates and a $Q$-dimensional vector of regression coefficients $\boldsymbol{\gamma}_k$ pertaining to the response-specific covariates.
According to (\ref{lin_pred1}), the effect of unit-specific covariates $\mathbf{x}_i$ is the same for all the $d$ response variables within a particular cluster $k$. The elements $\{\eta_{kij}\}$ of the linear predictor differ, with respect to $j$, only due to the effect of the response-specific covariates $\mathbf{w}_j$. We note that in the case $d=1$ and $K>1$, the model reverts back to a mixture of univariate GLMs \citep{wedel1995mixture, grun2008finite}; when $d>1$ but $K=1$, we obtain a multivariate GLM \citep{fahrmeir2013multivariate} and if both $K=1$ and $d=1$, we are back in a simple GLM \citep{mccullagh1989generalized}.

\subsection{Miro: mixture of regressions with overlap}
\label{subsec:tech}
Differently from the traditional setting descibed above, we are interested in defining a heterogeneous regression setting where units may  belong simultaneously to more than a single cluster. Using similar ideas to \cite{ranciati2017identifying}, we modify the hierarchical model in (\ref{hier1})  by relaxing conditions on the  allocation vectors $\mathbf{z}_{i}$ in order to allow for a multiple classification of the units. In particular, we will allow $\mathbf{z}_i\in \{0,1\}^K$.

If there are $K$ primary clusters, then there are $K^{\star}=2^K$ multiple cluster allocations. Each of these $K^\star$ allocations is a non-overlapping \emph{heir} cluster, which defines a new $K^{\star}$-dimensional allocation vector $\mathbf{z}^{\star}_i$ for each unit $i$. This vector satisfies $\sum_{h=1}^{K^{\star}} z^{\star}_{hi}=1$, and has a 1-to-1 correspondence with the original $\mathbf{z}_i$, which allocates units into the overlapping \emph{parent} clusters.
The $z^{\star}$ re-parametrization can now be used as the basis of a traditional hierarchical model, namely
\begin{eqnarray}\label{hier2}
\mathbf{z}^{\star}_i | \boldsymbol{\alpha}^{\star} &\sim& \text{Multinom}(\alpha^{\star}_1,\dots,\alpha^{\star}_{K^{\star}}),\\
\nonumber \mathbf{y}_i | \mathbf{z}^{\star}_i, \boldsymbol{\theta}^{\star}, \boldsymbol{\Sigma}^{\star} &\sim& \prod_{h=1}^{K^{\star}} \bigl[  \text{MVExpFam}\bigl(\boldsymbol{\theta}^{\star}_{hi};  \boldsymbol{\Sigma^{\star}_h} \bigl)\bigl]^{z^{\star}_{ih}}.
\end{eqnarray}
The key questions are (i) how to connect the new model parameters $\boldsymbol{\theta}^{\star}$ and $ \boldsymbol{\Sigma^{\star}}$ to the original parameters $\boldsymbol{\theta}$ and $ \boldsymbol{\Sigma}$; (ii) how to connect the new mixture parameters $\boldsymbol{\alpha}^{\star}$ to the old mixture parameters $\boldsymbol{\alpha}$ and (iii) how to interpret the resulting multiple allocation framework.


\paragraph{Overlap function: connecting heir parameters to parent parameters.}

In this paragraph we will describe a generic function for linking the new parameters $\boldsymbol{\theta}^{\star}$ and $ \boldsymbol{\Sigma^{\star}}$ with the original $\boldsymbol{\theta}$ and $ \boldsymbol{\Sigma}$, which we call the \emph{overlap function}. Various overlap functions are possible and each one of them  determines the way we can interpret and build the potential overlap between the components of the mixture. Each specific choice leads to different models that have their own computational and inferential considerations. Although the connecting function can be  directly at the level of the parameters $\boldsymbol{\theta}$, it is computationally more advantageous to define this  at the level of the linear predictors and regression coefficients. Furthermore, the overlap functions will be presented explicitly for the location parameters, but they are also natural choices for the nuisance parameters $ \boldsymbol{\Sigma}$.

Given the linear predictor $\eta_{kij}=\mu_k+ \boldsymbol{\beta}_k \mathbf{x}_i^{\intercal}  +  \boldsymbol{\gamma}_k \mathbf{w}_j^{\intercal},$ three natural choices to define the function $\eta_{ij}^\star = \psi( \boldsymbol{\eta_{.ij}}, \mathbf{z}_i)$ are the pointwise average, the minimum or the maximum of the linear predictors $\eta_{1ij}, \dots, \eta_{Kij}$ of the multiple allocations $\mathbf{z}_i$. In particular,
\begin{enumerate}
	\item \emph{minimum overlap function:}
	$ \psi_s( \boldsymbol{\eta_{.ij}}, \mathbf{z}_i)= \underset{k:z_{ik}=1}{\min}\{\eta_{kij}\}$,
	\item \emph{mean overlap function:}
	$ \psi_m( \boldsymbol{\eta_{.ij}}, \mathbf{z}_i)= \underset{k:z_{ik}=1}{\mbox{mean}} \{\eta_{kij}\},$
	\item \emph{maximum overlap function:}
	$ \psi_x( \boldsymbol{\eta_{.ij}}, \mathbf{z}_i)= \underset{k:z_{ik}=1}{\max} \{\eta_{kij}\}$,
\end{enumerate}
using {\bf s}mall, {\bf m}ean and ma{\bf x}imum as subscripts.
\begin{figure}[t!]
	\begin{center}
		\includegraphics[width=0.7\textwidth]{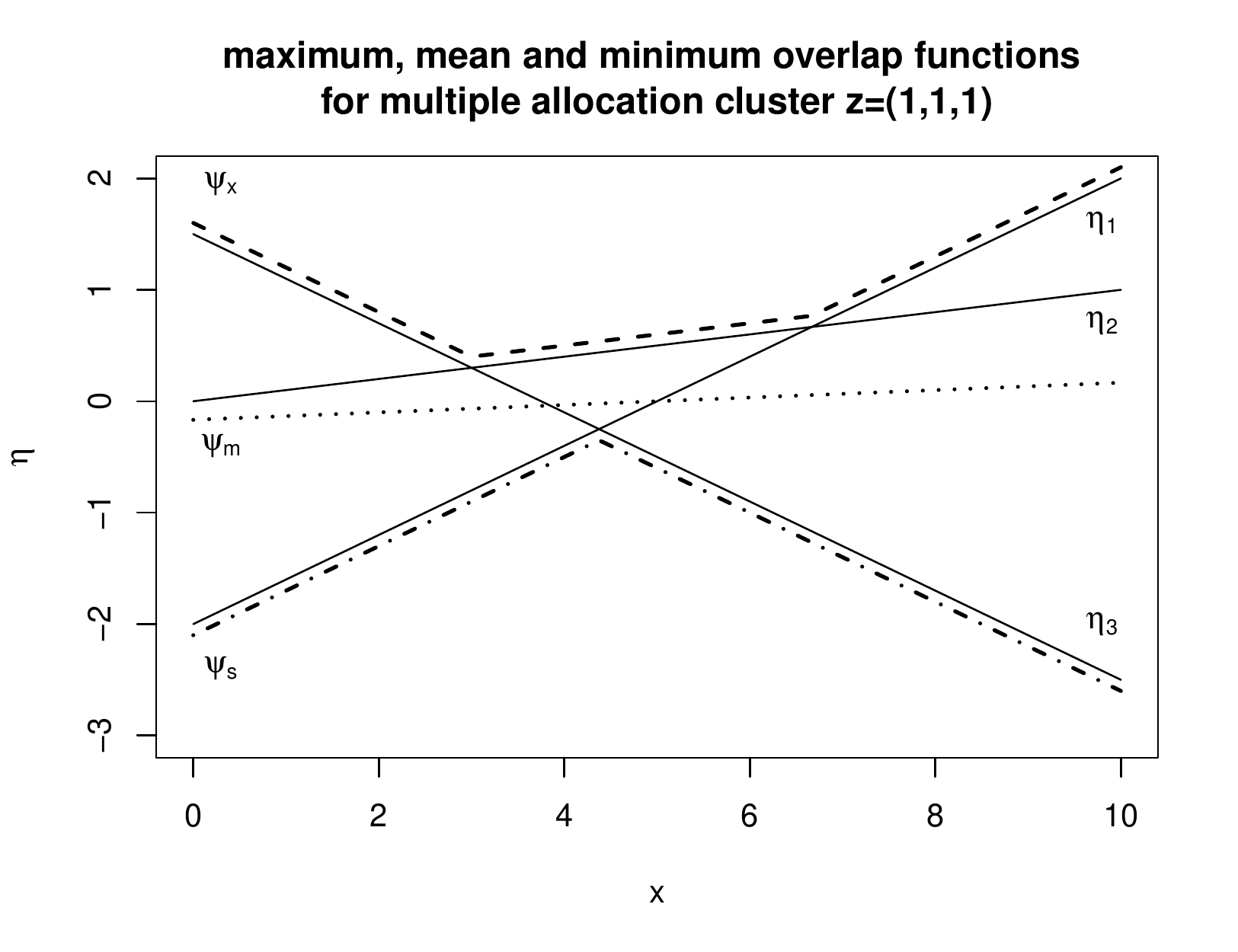}
		\caption{\label{fig:overlap} Example of the maximum, mean and minimum \emph{overlap functions} for a multiple allocation cluster $z=(1,1,1)$, for a single covariate $x$.}
	\end{center}
\end{figure}

Figure~\ref{fig:overlap} shows an example of the three overlap functions applied to a single covariate in a situation with three overlapping clusters. The mean overlap function is itself linear, whereas both the maximum and minimum overlap functions are only piecewise linear. In fact, the mean overlap function can be seen as an average of the intercepts and regression coefficients across those clusters, to which the unit is allocated,
\[ \psi_m(  \boldsymbol{\eta_{.ij}}; \mathbf{z}_i)
=\frac{\mathbf{z}_i  \boldsymbol{\mu}^{\intercal} }{{||\mathbf{z}_i||}_1}+\biggl(\frac{\mathbf{z}_i  \boldsymbol{B}}{{||\mathbf{z}_i||}_1}\biggl)\mathbf{x}_i^{\intercal} +\biggl( \frac{\mathbf{z}_i \boldsymbol{\Gamma}}{{||\mathbf{z}_i||}_1}\biggl)\mathbf{w}_j^{\intercal}, \]
where $\boldsymbol{\mu}=(\mu_1, \mu_2,\dots,\mu_K)$ is the vector containing all the $K$ intercepts, $ \boldsymbol{B}$ the $K \times L$ matrix whose rows are $\boldsymbol{\beta}_1,\dots,\boldsymbol{\beta}_K$, and $ \boldsymbol{\Gamma}$ the $K \times Q$ matrix with rows given by $\boldsymbol{\gamma}_1,\dots,\boldsymbol{\gamma}_K$. So, units in multiple allocation clusters have the average effects for each covariate with respect to their parent clusters.

For the case of being allocated to none of the primary parent clusters, $\mathbf{z}_i=0$, the overlap function requires a special definition. This definition depends on the individual situation. Sometimes it might be possible to define the overlap as an \emph{a priori} interpretable constant, whereas most other times it can be defined in a data driven way. In particular, one can use the overall minimum, $\psi( \boldsymbol{\eta_{.ij}},0)=\underset{x,w}{\min}\{\eta_{kij}(x,w)\}$, overall average, $\psi( \boldsymbol{\eta_{.ij}},0)=\underset{x,w}{\mbox{mean}}\{\eta_{kij}(x,w)\}$, or overall maximum, $\psi( \boldsymbol{\eta_{.ij}},0)=\underset{x,w}{\max}\{\eta_{kij}(x,w)\}$, on the observed range for $x$ and $w$. They are particularly recommended for the $\psi_x$, $\psi_m$ and $\psi_s$, respectively.

For the rest of this manuscript, we focus on the \emph{mean overlap function} $\psi_m$. This function offers computational advantages. Assuming that the $d$ response variables are independent, the mean overlap function translates the multivariate problem into a univariate GLM regression and allows estimations of all the coefficients of the $K$ clusters simultaneously. We give an intuition about the technical details of this implementation in the Appendix of the manuscript.

\paragraph{Connecting primary allocation parameters to heir parameters.} It is more complicated and restrictive to connect the allocation probabilities $\boldsymbol{\alpha}$ of the primary clusters $\boldsymbol{z}$ to the allocation probabilities $\boldsymbol{\alpha^\star}$ of the heir clusters $\boldsymbol{z^\star}$. In fact, the only meaningful, but restrictive choice is to use independent allocations, whereby the probability $\alpha^\star_{z^\star} = \prod_{k:z_k=1} \alpha_k$. In practical applications this is generally too restrictive and, therefore, the heir probabilities are estimated without any restrictions.

\subsection{Bayesian implementation}
\label{subsec:bayes}
We approach inference by following a Bayesian paradigm, which requires specification of prior distributions for all parameters in our model. First, we assume the prior cluster sizes $\boldsymbol{\alpha}^{\star}$ to come from a Dirichlet distribution with hyper-parameters $a_1,\ldots,a_{K^{\star}}.$ The intercepts $\mu_k$ and regression coefficients $(\boldsymbol{\beta}_k,\boldsymbol{\gamma}_k)$ are assumed to be a priori independent normally distributed centered at zero and with scalar variance parameters $(\sigma^2_\mu, \sigma^2_\beta, \sigma^2_\gamma)$ treated as hyper-parameters. The updated hierarchical formulation of the model is the following:
\begin{eqnarray}\label{hier3}
\boldsymbol{\alpha}^{\star} &\sim&\text{Dir}(a_1, \dots, a_{K^{\star}}),\\
\nonumber \mathbf{z}^{\star}_i | \boldsymbol{\alpha}^{\star} &\sim& \text{Multinom}(\alpha^{\star}_1,\dots,\alpha^{\star}_{K^{\star}}),\\
\nonumber \mu_k &\sim& \text{N}(0, \sigma^2_\mu), \\
\nonumber \boldsymbol{\gamma}_k &\sim& \text{N}_{Q}(\boldsymbol{0}_Q, \sigma^2_\gamma \, I_Q),  \\
\nonumber \boldsymbol{\beta}_k &\sim& \text{N}_{L}( \boldsymbol{0}_L, \sigma^2_\beta \, I_L),\\
\nonumber \mathbf{y}_i | \mathbf{z}^{\star}_i, \boldsymbol{\mu}, \boldsymbol{\beta}, \boldsymbol{\gamma}, \boldsymbol{\Sigma^\star} &\sim& \prod_{h=1}^{K^{\star}} \bigl[ \text{MVExpFam}\bigl(\boldsymbol{\theta}^{\star}_{hi}(\boldsymbol{\mu},\boldsymbol{\beta},\boldsymbol{\gamma}),\Sigma_h^\star\bigl)\bigl]^{z^{\star}_{ih}}.
\end{eqnarray}
Once one writes the complete joint posterior distribution of all the parameters and latent quantities in the model, an MCMC algorithm can be defined to sample from it. This will depend on the specific distributional choices, as we shall see with a specific example in Section~\ref{sec:probit}. In general, the pseudo-code for \texttt{miro} is an MCMC algorithm with Gibbs samplers, that iterates the following instructions across $t=1,\dots, T$ iterations:
\begin{enumerate}
 \item Use the full conditional of $\mathbf{z}^{\star}_{i}$, which is $$f(z^{\star}_{ih}=1 | y, \boldsymbol{\alpha}^{\star}, \boldsymbol{\beta},\boldsymbol{\gamma}, \boldsymbol{\mu}, \boldsymbol{\Sigma^\star})= \frac{\alpha^{\star}_h \cdot f(\mathbf{y}_i | z^{\star}_{ih}, \boldsymbol{\beta},\boldsymbol{\gamma}, \boldsymbol{\mu}, \boldsymbol{\Sigma^\star})}{\sum_{h'=1}^{K^{\star}}\alpha^{\star}_{h'} \cdot f(\mathbf{y}_i | z^{\star}_{h'i}, \boldsymbol{\beta},\boldsymbol{\gamma}, \boldsymbol{\mu}, \boldsymbol{\Sigma^\star})}$$
 to sample the allocation vectors $\{ \mathbf{z}^{\star}_{i} \}_{i=1,\dots,n},$ and then compute the updated cluster sizes $n_{h}=\sum_{i=1}^n z^{\star}_{ih};$
 \item Sample new values for $\{\alpha^{\star}_h\}$ from their full conditional $$\boldsymbol{\alpha}^{\star} | y, \mathbf{z}, \mathbf{a} \sim \text{Dir}(a_1+n_1, \dots, a_h+n_h,\dots,a_{K^{\star}}+n_{K^{\star}});$$
 \item Sample intercepts and regression coefficients $(\boldsymbol{\mu}, \boldsymbol{B}=\{\boldsymbol{\beta}\}, \boldsymbol{\Gamma}=\{\boldsymbol{\gamma} \})$, by using the stacked design matrices $\boldsymbol{X}$ and $\boldsymbol{W}$, from their full conditional distributions. The instructions inside this step are dependent on distributional assumptions for the data matrix $y$; we will see a specific example in Section \ref{sec:probit};
\end{enumerate}

\subsection{Model selection, posterior allocation, and label switching}
In the context of our approach, model selection is equivalent to choosing the number of primary clusters $K$. A fully Bayesian specification would prescribe a prior on this quantity and, due to the nature of the model, would require the implementation of a trans-dimensional MCMC version of our algorithm, such as a reversible jump MCMC \citep{green1995reversible}. For two reasons we avoid this approach. First, we expect the `true' number of primary clusters $K$ to be small in practice, as even $K=4$ can accommodate for up to $K^{\star}=16$ clusters. Second, sampling $K$ from its prior distribution might lead to computational issues. Given that $K^{\star}$ scales exponentially with $K$, the support of the prior distribution would have to be rather narrow to avoid unfeasible values of $K^{\star}$, which defeats the main purpose of using a distribution for $K$.

On the basis of these considerations, we opt for a more heuristic model selection approach, in which we fit our model for different values of $K$ and then select the optimal one through an information criterion. In particular, we rely on the
BIC-MCMC \citep{fruhwirth2011label}, recently employed to select infinite mixtures of infinite factor analyzers models \citep{murphy2017infinite}.
The BIC-MCMC criterion typically encourages the selection of parsimonious models, and it is defined as $$\text{BIC-MCMC}= -2l_{\text{max}} + \log(n\cdot d) \cdot p_{\theta}$$where $l_{\text{max}}$ is the maximum value of the log-likelihood across the MCMC chain (after burn-in) and $p_{\theta}$ is the number of parameters in the model.

After the choice of $K$ and, implicitly, ${K}^{\star}$, units are allocated into clusters according to their average posterior probabilities and using the Maximum-A-Posteriori (MAP) rule. In particular, unit $i$ will be assigned to the cluster $h$ that attains the highest value for $$\bar{\text{P}}(\mathbf{z}^{\star}_i=h | y, \boldsymbol{\alpha}^{\star}, \boldsymbol{\theta}, \boldsymbol{\Sigma})= \frac{1}{T} \sum_{t=1}^{T}\text{P}(\mathbf{z}^{\star}_i=h | y,\boldsymbol{\alpha}^{\star}, \boldsymbol{\theta}, \boldsymbol{\Sigma}),$$ computed after the initial burn-in window.

A well-known problem with mixture models in the Bayesian paradigm is the label switching phenomenon \citep{celeux1998bayesian,stephens2000dealing,sperrin2010probabilistic}. Although, in theory, a desirable property of the formulation, as it allows the MCMC chain to visit all the modes of the target distribution, the label switching arises from the invariance property of the likelihood with respect to the order of the cluster labels. From a practical point of view, this is reflected in unwanted difficulties while summarizing posterior quantities, i.e., posterior means, posterior standard deviations, etc, for some parameters of interest.
To tackle this issue, we reorder the MCMC output of the algorithm through the geometrically-based Pivotal Reordering Algorithm (PRA) \citep{marin2005bayesian,marin2007bayesian}, available as a function in the \texttt{R} package \texttt{label.switching} \citep{labswitch}.
The procedure needs a pivotal labelling, which we select according to the strategy proposed in \citep{carmona2018model}. In particular, we first compute the matrix of co-occurences $C^{(t)}$ at each iteration $t=1,\dots,T$ of the MCMC (after burn-in). This is an $n \times n$ matrix where a generic element $c_{ij}$ is equal to one if unit $i$ is in the same cluster of unit $j$ for that iteration, and zero otherwise. Then, an average of these matrices is computed, denoted by $\bar{C}$. Finally, we select the labelling of iteration $t_{\text{min}}$ as our pivotal quantity, where $$t_{\text{min}}= \argmin_{t}\bigl\{ \big[C^{(t)}-\bar{C}\big]^2\bigl\}.$$ Once samples have been relabelled according to the algorithm, we can compute posterior quantities of interest.

\section{Two-mode binary networks with covariates}
\label{sec:probit}
Two-mode networks \citep[Chapter 8]{wasserman1994social}, also known as bipartite or affiliation networks, are networks consisting of two types of nodes, where links can occur only between nodes of different type. An example of a two-node network is a group of \emph{actors} attending or not attending a set of \emph{events}. From an agent-based point of view, a two-mode network can be seen as a collection of $d$-dimensional multivariate binary response variables for $n$ actors. For each of the actors, the response vector indicates whether he or she attended or did not attent each of the $d$ events.

One of the interests in two-mode networks is detecting clusters of actors that represent organizational structures inside the network itself. Recent studies have shown how allowing for overlap can improve the characterization of the clusters as well as leading to model parsimony \citep{ranciati2017identifying}. With additional data often available, both at the level of actors (units) and events (responses), and with a natural expectation of clusters to be potentially overlapping, we present this setting as a prime example of our mixture of multivariate GLMs (\texttt{miro}) approach presented in Section~\ref{sec:multiGLMs}.

We organize our data in an $n \times d$ matrix $y$ of observations $y_{ij}$, recording attendances of $i=1,\dots,n$ units or \emph{actors} to $j=1,\dots,d$ \emph{events}. Each $y_{ij}$ is a realization from a binary random variable, where $y_{ij}=1$ indicates that individual $i$ attended event $j$, and zero otherwise. We assume $\{y_{ij}\}$ to be (conditionally) independent for all $i,j$. 
The hierarchical formulation of the model is therefore
\vspace{-2mm}
\begin{eqnarray}
 \nonumber \boldsymbol{\alpha}^{\star} &\sim&\text{Dir}(a_1, \dots, a_{K^{\star}}),\\
 \nonumber \mathbf{z}^{\star}_i | \boldsymbol{\alpha}^{\star} &\sim& \text{Multinom}(\alpha^{\star}_1,\dots,\alpha^{\star}_{K^{\star}}),\\
 \nonumber \mathbf{y}_i | \mathbf{z}^{\star}_i,\boldsymbol{\mu}, \boldsymbol{\beta}, \boldsymbol{\gamma} &\sim&  \prod_{h=1}^{K^{\star}} \prod_{j=1}^d \biggl[ \bigl(\pi^{\star}_{hij}\bigl)^{y_{ij}}\bigl(1-\pi^{\star}_{hij}\big)^{1-y_{ij}} \biggl]^{z^{\star}_{ih}},
\end{eqnarray}
where probabilities of attendance $\{\pi^{\star}_{kij}\}$ play the role of $\{ \theta_{hi}^{\star}\}$ in   (\ref{hier3}).
A natural choice for overlap function of the zero cluster, $\mathbf{z}_i=(0,\dots,0)$, is to set $\pi^\star =0$  and use it to allocate units for which we record no attendances.
Moreover, for the binomial setting, we note that there are no additional dispersion parameters $\boldsymbol{\Sigma}$.

In this two-mode binary network setting, covariates in $X$ could be characteristic related to an actor, such as gender, age, etc, and those in $W$ could be features of an event, i.e., type of event, date, duration, and so forth.
We bridge the linear predictors and their corresponding probabilities of attendance via $g(\cdot)$, which we choose to be the Gaussian cumulative distribution function $\pi^{\star}_{hij}=\Phi\bigl(\eta^{\star}_{hij}\bigl).$ This choice of a link function induces a probit regression model formulation, which, combined with the mean overlap function, allows one to sample the parameters of interest, $\{\boldsymbol{\mu},\boldsymbol{\beta},\boldsymbol{\gamma}\}$, directly from a single probit regression model instead of working separately for each cluster $k=1,\dots,K.$ To be more specific, we can write the likelihood as
$$L_{y,z}\propto \prod_{i=1}^{n\cdot d} \bigl[\Phi(\eta_{ij}^\star)\bigl]^{y_{ij}}\bigl[1-\Phi(\eta_{ij}^\star)\bigl]^{1-{y}_{ij}}$$ and pair this with the same prior distributions shown in (\ref{hier3}), in order to derive the full joint posterior of the parameters in our model.

For inferential purposes, we employ the Bayesian probit regression framework proposed by \cite{holmes2006bayesian} and we implement the MCMC scheme previously described in Section \ref{subsec:bayes}. More specifically, in Step 3 of the pseudo-code, the set of regression coefficients, $\boldsymbol{\mu}$, $B$, and $\Gamma$, is sampled together in a single probit regression step. We adopt the framework suggested by \cite{holmes2006bayesian}, where a random latent utility $r$ is introduced, such that $y_{ij}=1$ if $r_{ij}>0$, otherwise ${y}_{ij}=0,$ for $i=,\ldots,n$ and $j=1,\ldots d.$ Then utility is defined as $r_{ij}=\eta_{ij}^\star+\epsilon_{ij}$ with $\epsilon_{ij} \sim \mathcal{N}(0,1)$ and the linear predictor $\eta_{ij}^\star$ is the same of a probit regression on the original ${y}_{ij}.$ The approach leads to the following update mechanism for the regression coefficients $\boldsymbol{\theta}^{\intercal}=[ {\boldsymbol{\beta}} \,\,\, {\boldsymbol{\gamma}} ]$ in $\eta_{ij}^\star:$
\begin{enumerate}
\item[i)]  $D \leftarrow 0;$
\item[ii)]  for $i=1,\dots, n$ and $j=1,\ldots, d$
\begin{eqnarray}
\nonumber m_{ij} &\leftarrow& A_{ij} \, \boldsymbol{\theta}^{(t-1)}\\
\nonumber r_{ij} &\leftarrow& \text{\texttt{truncNorm}}({y}_{ij}; m_{ij}, 1)\\
\nonumber D &\leftarrow& D+r_{ij} S_{ij}
\end{eqnarray} with $\text{truncNorm}(\cdot)$ denoting sample values from a truncated Normal distribution;

\item[iii)] $C \leftarrow \text{N}(\boldsymbol{0}, I);$
\item[iv)] $\boldsymbol{\theta}^{(t)} \leftarrow D+\text{Chol}(V)^{\top}C$
\end{enumerate}
where $V$ is the prior block-covariance matrix of the coefficients in $\boldsymbol{\theta}$; $A$ is the full design matrix obtained by stacking all the columns of both ${X}$ and ${W}$, $S=V\,A^{\top}$ and $\text{Chol}(\cdot)$ extracts the lower-triangular matrix from the Cholesky decomposition.

\section{Simulation study}
\label{sec:simul}


We investigate the performance of our model under two different data generating processes: (i) data coming from our model, specifically in the form presented in Section \ref{sec:probit}; (ii) data obtained from a mixture of non-overlapping components, with covariates and a logit link function. For each case, we simulate 25 independent datasets, and we average the results across the replicates. The performance is measured via: (i) the misclassification error rate (\emph{MER}), which is the fraction of wrongly allocated units with respect to the true labeling, and (ii) the Adjusted Rand Index (\emph{ARI}), that measures how much two labelings (true one and sampled one) agree with each other. Benchmarks values for \emph{MER} and \emph{ARI} are, respectively, 0.685 and 0: the first is computed as $\sum_k[\alpha_k(1-\alpha_k)]$, where $\boldsymbol{\alpha}=(0.10,0.45,0.25,0.20)$ is the chosen vector of cluster sizes, thus it is the probability of a wrong allocation under random assignments of units to clusters; the second is by definition the rand index obtained under a random allocation.

\subsection{Synthetic data from \texttt{miro} model}
\label{subsec:sim}
Data are simulated from $K=2$ overlapping clusters, using the hierarchical formulation of the \texttt{miro} model as the data generating process, namely a mixture model with overlapping components and a probit regression formulation. We consider 5 scenarios in total, according to the type of covariates considered, and by varying either the sample size $n$ or the number of events $d$. 
In particular, we consider:
\begin{itemize}
\item Settings with \emph{Actor} covariates only:  sample size $n=50$ actors, $d=\{5,20\}$ events, $L=1$ continuous covariate $X$ sampled from a Standard Normal distribution;
\item Settings with \emph{Event} covariates only: sample size $n=\{50,150\}$ actors, $d=15$ events, $Q=2$ binary covariates $(W_1,W_2)$ from one categorical covariate with three levels;
\item Setting with \emph{Actor} and \emph{Event} covariates: sample size $n=250$ actors, $d=21$ events, $L=1$ continuous covariate $X$ sampled from a Standard Normal distribution, $Q=2$ binary covariates $(W_1,W_2)$ from one categorical covariate with three levels.
\end{itemize}
The results are reported in Table \ref{res_dgp_manet}. With the exception of the scenario with the lowest sample size and number of events,  $n=50$ and $d=5$, respectively, we are able to obtain appreciable values for both \emph{MER} and \emph{ARI}.  This is indeed expected, given that we are simulating from the same model which we use for inference. However, we further notice that not only sample size but also increasing the number of events leads to better performances. In particular, increasing the number of events $d$ has a positive effect for the setting with only \emph{actor} covariates: this is due to the fact that, having more attendances is analogous to having more time points in a repeated measures model framework. A similar argument can be made for the effect of sample size $n$ on the performances in scenarios where there are only \emph{event} covariates.
\begin{table}
\centering\begin{tabular}{c|c|c|c}
\hline
Type of covariates & Sample size and \# of events & Misclass. Err. Rate & Adj. Rand Index \\
\hline
\emph{actor} & $n=50, \,\,d=5$ & 42.08 & 19.52 \\
\emph{actor} & $n=50, \,\,d=15$ & 14.72 & 68.51 \\
\hline
\emph{event} & $n=50, \,\,d=15$ & 15.60 & 67.62 \\
\emph{event} & $n=150, \,\,d=15$ & 13.73 & 70.61 \\
\hline
\emph{actor}, \emph{event} & $n=250, \,\,d=21$ & 18.00 & 66.00 \\
\hline
\end{tabular}\caption{Simulation study with data generated from \texttt{miro}: misclassification error rate and adjusted rand index are averaged across the 25 replicated datasets and reported as percentages.\label{res_dgp_manet}}
\end{table}

\subsection{Synthetic data from misspecified model}
\label{subsec:mispecsim}
We now simulate data from $K=4$ non-overlapping groups via a mixture model of binary regression models, where probabilities $\{\pi_{kij}\}$ are computed as a logit transformation of the linear predictor $$\eta_{kij}=\mu_k+\beta_{k1}x_{i1}+\gamma_{k1}w_{j1}.$$ Here, we simulate $n=300$ units and $d=20$ events. This data generating process differs from $\texttt{miro}$ due to: (i) logit instead of probit link function; (ii) units belong only to one component at a time, as in a conventional mixture model. As covariates, we use: $L=1$ continuous \emph{actor} covariate $X$ sampled from a standard Normal distribution; $Q=1$ binary \emph{event} covariate $W.$ We perform inference using four competing models: (i) \texttt{mixtbern}, a conventional mixture of Bernoulli distributions; (ii) \texttt{manet}, a mixture of Bernoulli distributions with overlapping clusters \citep{ranciati2017identifying}; (iii) \texttt{mixtprobit}, a classical mixture of probit regression models; (iv) \texttt{miro}, our proposed model. For \texttt{manet} and \texttt{miro}, prior on the cluster sizes $\text{P}(\alpha_1^{\star},\alpha_2^{\star},\dots,\alpha_h^{\star},\dots,\alpha_{K^{\star}}^{\star})=\text{Dir}(a_1,a_2,\dots, a_h,\dots,a_{K^{\star}})$ are set to have the following hyper-parameters: $a_h=K^{\star}$ if $\sum_{h=1}^{K^{\star}}u_h=1$, otherwise $a_h=1$.

The results are reported in Table \ref{res_mispec_nocor_d20_newprior}.
\begin{table}[h!]
\centering\begin{tabular}{r|c|c}
\hline
Model & Misclass. Err. Rate & Adj. Rand Index \\
\hline
\texttt{mixtbern}, $K=4$ & 44.69 & 16.11 \\
\hline
\texttt{manet}, $K=2$ & 46.61 & 20.61 \\
\texttt{manet}, $K=3$ & 44.92 & 16.62 \\
\texttt{manet}, $K=4$ & 48.87 & 15.35  \\
\hline
\texttt{mixtprobit}, $K=2$ & 41.88 & 29.74 \\
\texttt{mixtprobit}, $K=3$ & 27.09 & 49.35 \\
\texttt{mixtprobit}, $K=4$ & 25.25 & 50.86 \\
\hline
\texttt{miro}, $K=2$ & 41.65 & 25.67 \\
\texttt{miro}, $K=3$ & 26.37 & 50.26 \\
\texttt{miro}, $K=4$ & 29.01 & 45.37 \\
\hline
\end{tabular}\caption{Misclassification error rate and Adjusted Rand index, averaged across the replicated datasets and reported as percentages; $K=4$ non-overlapping components, logit link function, uncorrelated data, $10$ replicated datasets, $n=300$ units, $d=20$ events.\label{res_mispec_nocor_d20_newprior}}
\end{table}
Performances degrade in this setting with respect to the previous section, due to the misspecification of the models we fit. However, the results for \texttt{miro} with $K=3$ and $K=4$ are less affected than those for the two competing models, \texttt{mixtbern} and \texttt{manet}. As expected, being the closest to the data generating process,  \texttt{mixtprobit} performs on par or slightly better than \texttt{miro} in terms of ARI and MER.

\section{Agreement and polarization in  U.S. Supreme Court}
\label{sec:supreme}
We illustrate the performance of our proposed method on the 26 ``important decisions'' handed down by the US Supreme Court in their 2000-2001 term with the aim of clustering its nine justices. Originally described in \cite{greenhouse2001year}, these data were analyzed in \cite{doreian2003structures} and then further explored in \cite{doreian2004generalized}. The $n=9$ \emph{actors} in our data are justices Breyer, Ginsburg, Souter, Stevens, O'Connor, Kennedy, Rehnquist, Scalia and Thomas, whereas the \emph{decisions} considered are $d=26$. According to \cite{greenhouse2001year}, the decisions can be categorized into 7 main topics, which can be used as a categorical covariate $W$ with the following levels: ``Presidential Election'', ``Criminal law'',
``Federal authority'', ``Civil rights'', ``Immigration law'', ``Speech and Press'', ``Labor and Properties''. Each observation is coded as $y_{ij}=1$ when  justice $i$ was part of the majority decision $j$, while $y_{ij}=0$ stands for the situation where the justice voted with the minority.

On these data we apply four clustering algorithms: \texttt{mixtbern}, \texttt{manet}, \texttt{mixtprobit} and \texttt{miro}. For all of them, we opt for 10,000 MCMC iterations with a generous 5,000 burn-in window. Overall model fit comparison is done quantitatively in terms of BIC-MCMC and qualitatively using the clustering output. In Table~\ref{tab_supreme} we report the selected number of clusters $K$ according to the BIC-MCMC value. We also provide a proxy for the complexity of the models by reporting the number of parameters.

We gather from the summary that \texttt{miro} stands out as the best model according to the BIC-MCMC. Moreover, \texttt{manet} and \texttt{miro} produce the same classification of the justices, which suggests that no real better fit is provided by \texttt{manet} at the expense of increasing model complexity. In particular, justices Breyer, Ginsburg, Souter, and Stevens are allocated into primary cluster $(1,0)$, whereas Rehnquist, Scalia, and Thomas are grouped together in the second primary cluster (0,1). Appropriately,  O'Connor and Kennedy are allocated into the multiple allocation cluster $(1,1)$.
It has been well-documented that Kennedy and O'Connor constituted the swing vote in the Supreme Court \citep{toobin2008nine}.

On the other hand, \texttt{mixtbern} identifies three separate clusters, where Kennedy is put together with Rehnquist, Scalia, and Thomas, while O'Connor is allocated alone into a third group. Finally, BIC-MCMC suggests a value of $K=3$ in the case of \texttt{mixtprobit}, although the posterior classification of the units leaves one cluster empty, with Breyer, Ginsburg, Souter and Stevens in one cluster and O'Connor, Kennedy, Rehnquist, Scalia and Thomas in the other. These clusterings seem less meaningful.

\begin{table}[tb]
\centering\begin{tabular}{c|c|c|c}
\hline
Model & $K$, \# of cluster & BIC-MCMC & \# of parameters\\
\hline
\texttt{mixtbern} & $3$ &  $571.06$ & 81 \\
\texttt{manet} & $2$ & $438.73$ & 56 \\
\texttt{mixtprobit} & $3$ & $368.15$ & 24  \\
\texttt{miro} & $2$ & $\mathbf{332.28}$ & 18 \\
\hline
\end{tabular}\caption{\label{tab_supreme} Model selection criterion reported for the three competing algorithms; $K$ is the number of cluster selected for each model.}
\end{table}

\begin{figure}[t!]
	\begin{center}
		\includegraphics[width=0.95\textwidth]{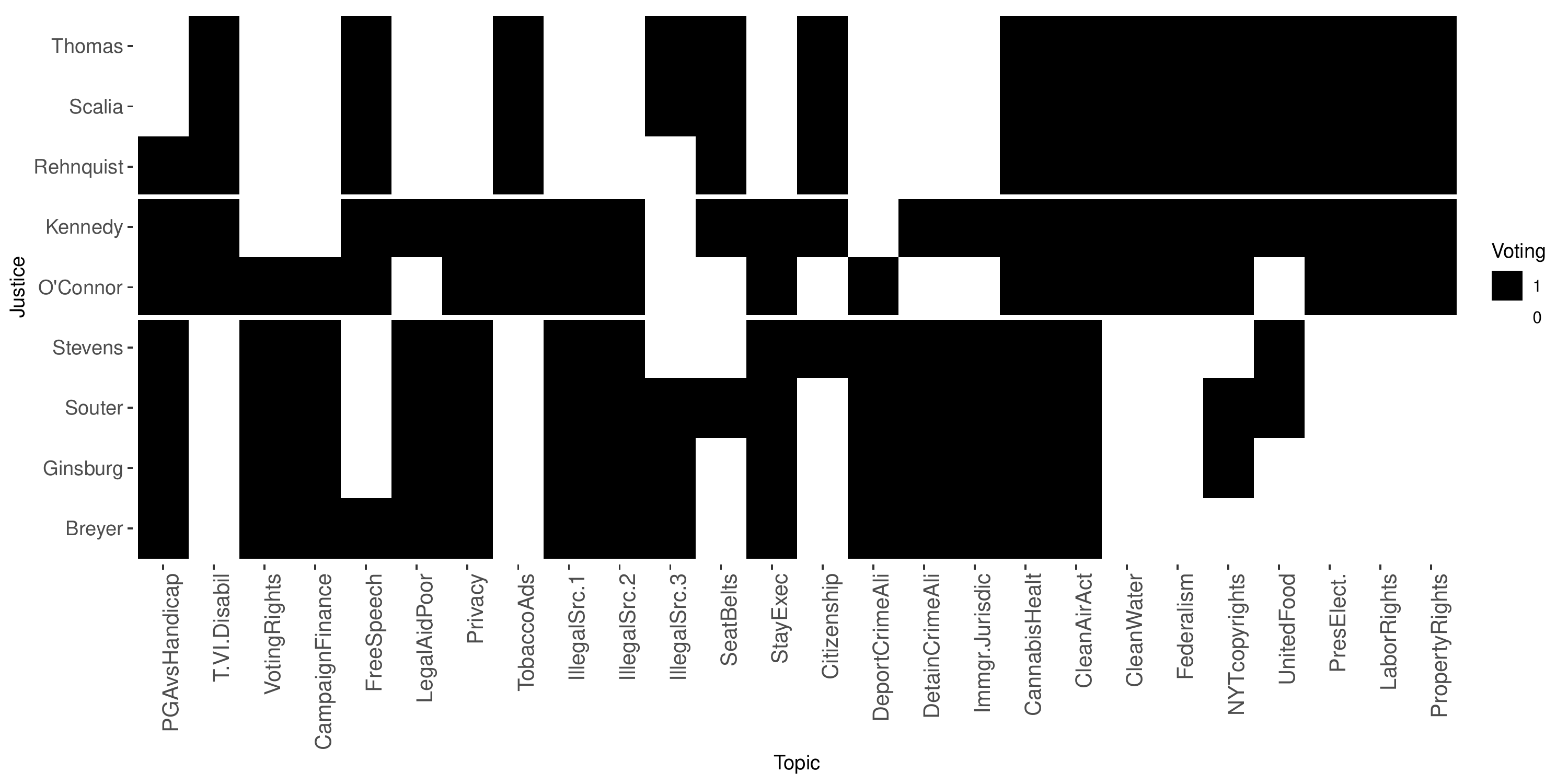}
		\includegraphics[width=0.95\textwidth]{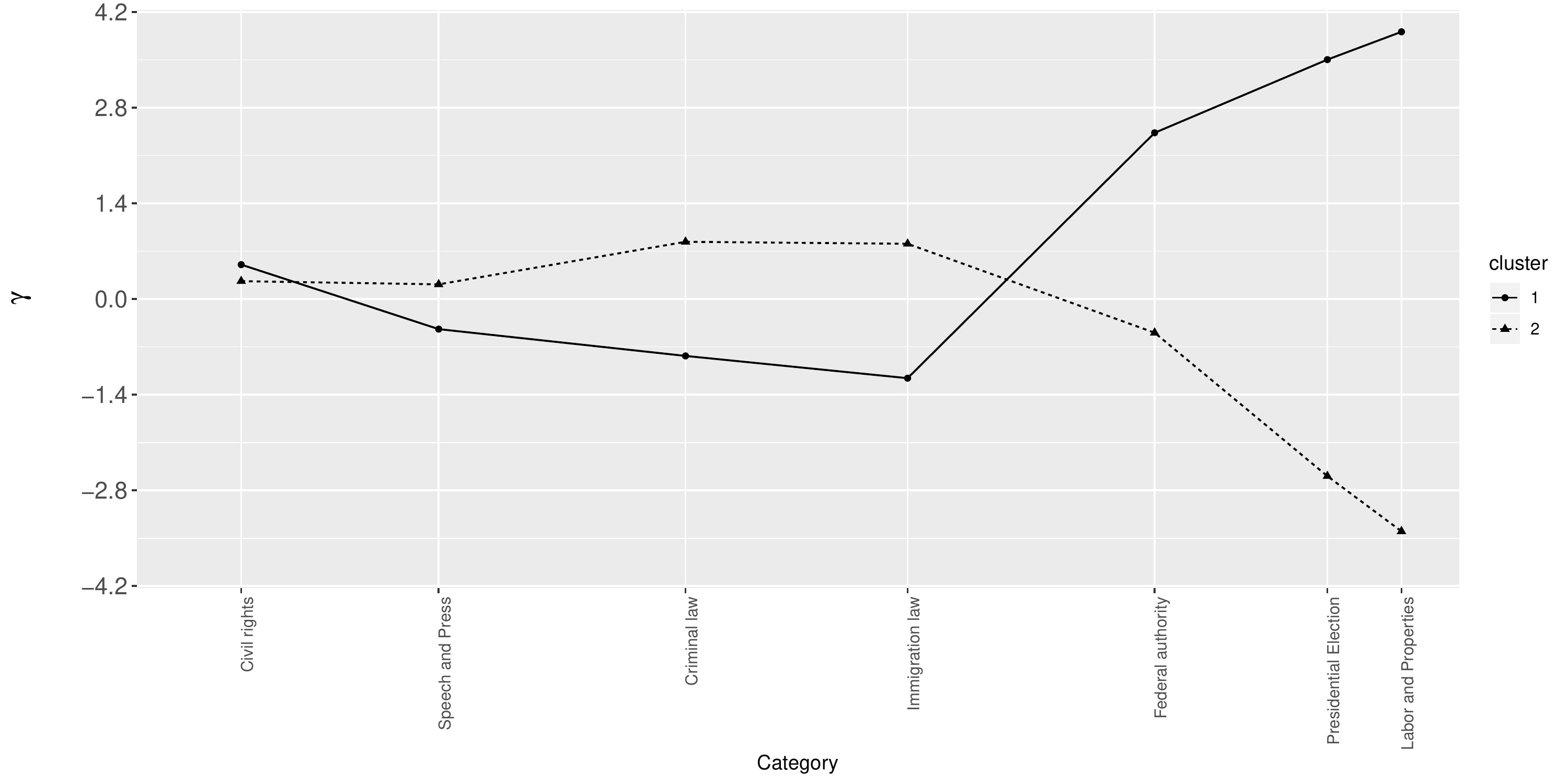}
		\caption{\label{plot_supreme} Results for \texttt{miro} with $K=2$; (\emph{top plot}) black tiles represent observations equal to 1, while empty tiles code for 0; row names are the justices, while column names indicate decisions: tiles are horizontally separated by white lines according to clustering, resulting in three sections; (\emph{bottom plot}) regression coefficients associated to each category for the covariate ``type of decision'', with solid black line for primary cluster $k=1$ and the dotted black line for the other primary cluster $k=2$.}
	\end{center}
\end{figure}

Unlike any of the other methods, \texttt{miro} is able to make use of additional covariate information to aid the clustering. Figure~\ref{plot_supreme} visualizes the results for \texttt{miro} in terms of clustered data and posterior means of regression coefficients for the two primary clusters.
Coherently with the allocations of the model, there are some decision types that
better discriminate between the voting behaviour of the two primary clusters. In particular, those decisions belonging to categories ``Federal Authority'',
``Presidential Election'' and ``Labor and Properties'' on the right side of bottom
plot in Figure~\ref{plot_supreme} clearly discriminate the liberal judges in cluster 2 from the conservative judges in cluster 1. On the other four decision categories, the 9 justices are in much closer agreement.

\section{Conclusions}
\label{sec:concl}

In this manuscript we have presented an approach to perform model-based clustering on multivariate data, via a mixture of generalized linear models that allows for units to be allocated in more than one cluster, while incorporating additional information in the form of covariates. The proposed method has the advantage of allowing the user to define, from a modeling perspective, how the multiple allocation clusters are related to the main parameters of the GLM, and, in particular, how to combine the regression coefficients in order to have results with a high degree of interpretability as well as to aid the identification of clusters. The multivariate GLM method presented is quite general, both in terms of the type of data that can be modeled and the Bayesian inference procedure that can be applied.

We illustrated the method by Bayesian implementation of a probit regression model for two-mode network data, which can be seen as collections of binary observations for different units attending or not-attending events, or in the example presented, for different US supreme court justices voting with the majority or with the minority on 26 important decision in 2000-2001. A simulation study provided encouraging results with respect to the performance of the method  related to the identification of clusters, even in situations where the simulated environment is not the same as the fitted model. The comparison was favorable also against some close competitors, such as mixtures of GLMs without overlapping clusters or mixtures with overlapping clusters but without covariates. Finally, we adopted our proposed methodology to the voting records of the US supreme court and identified interesting clusters of voting behaviour for its 9 justices in the 2000-2001 term. The data and R code for the analysis can be found in \url{https://github.com/savranciati/miro}.

When considering possible future extensions of the method, the definition of the linear predictor could be further enriched by adding, for example, random effects for knowing grouping structures, or specifying regression coefficients that vary for each possible combination of cluster $k$, unit $i$, and variable $j$. However, this will have the drawback of significantly increasing the number of parameters to be inferred. Also, further extensions could revolve around the idea of introducing dependence between the response variables. In this direction, \cite{karlis2010} adopted Copula models in the context of multivariate GLMs, while other authors explored the case of mixtures of bivariate Poisson GLMs \citep{BermudezKarlis2012}.

\bibliographystyle{jasa3}
\bibliography{bib_overlapping_GLMs}

\appendix

\section{Technical details on the \emph{mean overlap function} implementation}

First, we let $\tilde{\mathbf{y}}^{\intercal}$ be an $\tilde{n} \times 1$ vector obtained by stacking columns of data matrix $y$, with $\tilde{n}=n\cdot d$.  Furthermore, we stack together the cluster-specific vector of intercepts and regression coefficients for individual covariates into a vector $\tilde{\boldsymbol{\beta}}=[\mu_1 \,\,\, \boldsymbol{\beta}_1 \,\,\, \mu_2 \,\,\, \boldsymbol{\beta}_2 \,\,\, \dots \,\,\, \mu_K \,\,\, \boldsymbol{\beta}_K]$; we do the same with response-specific covariates coefficients as a vector $\tilde{\boldsymbol{\gamma}}=[\boldsymbol{\gamma}_1 \,\,\, \,\,\, \boldsymbol{\gamma}_2 \,\,\, \dots \,\,\, \boldsymbol{\gamma}_K]$ . Accordingly, we can define a GLM simultaneously for all components, where
\begin{eqnarray}
\nonumber \tilde{y}_i | \tilde{\boldsymbol{\beta}},  \tilde{\boldsymbol{\gamma}}, \mathbf{z}_i  \sim \text{MVExpFam}\bigl(g(\tilde{\eta_i})\bigl)
\end{eqnarray}
with $\tilde{\eta}_i$ being an element of $\tilde{\boldsymbol{\eta}}=\mathbf{\tilde{X}} \tilde{\boldsymbol{\beta}}^{\intercal} + \mathbf{\tilde{W}} \tilde{\boldsymbol{\gamma}}^{\intercal}.$ The design matrices $\mathbf{\tilde{X}}$ and $\mathbf{\tilde{W}} $ are built by filtering the corresponding matrices of covariates' values $\mathbf{X}$ and $\mathbf{W}$ through the allocation of each unit, and building a block structure to reflect the possible configurations of $\mathbf{z}_i$. In particular, we collect in a matrix $\mathbf{X}_{[h]}$ the predictors' recorded values for the $n_h=\sum_{i=1}^{n} z^{\star}_{hi}$ units allocated into cluster $h$, and we stack them vertically $d$ times; also, we append a column unitary vector of length $n_h$ to account for the intercept. The matrix $\mathbf{W}_{[h]}$ is simply built by stacking $\mathbf{W}$ exactly $n_h$ times, in order to have conforming dimensions. All the resulting matrices have $n_{h} \times d$ rows and number of columns, respectively, $(L+1)$ and $Q.$ This process is repeated for $h=1,\dots,K^{\star}$. Finally, the block structures of $\mathbf{\tilde{X}}$ and $\mathbf{\tilde{W}}$ are defined such that
\begin{equation}\nonumber
\mathbf{\tilde{X}}=\left[ \begin{array}{c}
\mathbf{u}_1 \otimes  \mathbf{X}_{[1]}  \\[6px]
\vdots  \\[6px]
\mathbf{u}_h  \otimes \mathbf{X}_{[h]} \\[6px]
\vdots  \\[6px]
\mathbf{u}_{K^{\star}} \otimes  \mathbf{X}_{[K^{\star}]}  \\
\end{array}\right] \,\,\,\,\,\,\,\,\,\, \,\,\,\,\,\,\,\,\,\, \mathbf{\tilde{W}}=\left[ \begin{array}{c}
\mathbf{u}_1 \otimes \mathbf{W}_{[1]}  \\[6px]
\vdots  \\[6px]
\mathbf{u}_h \otimes \mathbf{W}_{[h]} \\[6px]
\vdots  \\[6px]
\mathbf{u}_{K^{\star}} \otimes \mathbf{W}_{[K^{\star}]}  \\
\end{array}\right]
\end{equation}
where $\mathbf{u}_h$ is the $h$-th row of $U$, and $\otimes$ is the Kronecker product. The final design matrices $\mathbf{\tilde{X}}$ and $\mathbf{\tilde{W}} $ have $\tilde{n}=n \cdot d$ rows and, respectively, $(L +1) \times K$ and $Q \times K$ columns. The sub-matrices involving $h=1$ are not used to sample the regression coefficients. The matrix $U$ is defined to contain all the possible configurations of 1s and 0s of length $K$, so that we have $z^{\star}_{hi}=\indicator{\mathbf{u}_h=\mathbf{z}_i}$ with $\mathbf{u}_h$ denoting the $h$-th row of $U$ and $\indicator{\cdot}$ the indicator function.

For example, when $K=2$ and only actor-specific covariates are considered, the relevant quantities are

\begin{eqnarray}
\nonumber U= \begin{pmatrix}
0 & 0 \\
1 & 0 \\
0 & 1 \\
1 & 1 \\
\end{pmatrix} = \left[ \begin{array}{c}
\mathbf{u}_1  \\[6px]
\mathbf{u}_2 \\[6px]
\mathbf{u}_3 \\[6px]
\mathbf{u}_4 \\[6px]
\end{array}\right]
\,\,\,\,\,\,\,\,\,\, \,\,\,\,\,\,\,\,\,\,
&\mathbf{\tilde{X}}=\left[ \begin{array}{cc}
\mathbf{0}_{1+L} & \mathbf{0}_{1+L}  \\[6px]
\mathbf{X}_{[2]} & \mathbf{0}_{1+L} \\[6px]
\mathbf{0}_{1+L} & \mathbf{X}_{[3]} \\[6px]
\frac{1}{2}\mathbf{X}_{[4]} & \frac{1}{2}\mathbf{X}_{[4]} \\
\end{array}\right] \,\,\,\,\,\,\,\,\,\, \,\,\,\,\,\,\,\,\,\,  &\tilde{\boldsymbol{\beta}}^{\intercal}=\left[\begin{array}{c} \mu_1 \\[3px] \boldsymbol{\beta}_{[1]} \\[6px] \mu_2 \\[6px] \boldsymbol{\beta}_{[2]}\end{array}\right]
\end{eqnarray}
and each unit $i$ may be assigned:
\begin{itemize}
\item to none of the two clusters, $\mathbf{z}_i=\mathbf{u}_1=(0,0)$, corresponding to $\mathbf{z}^{\star}_i=(1,0,0,0)$;
\item only to the first \emph{parent} cluster, $\mathbf{z}_i=\mathbf{u}_2=(1,0)$, corresponding to $\mathbf{z}^{\star}_i=(0,1,0,0)$;
\item only to the second \emph{parent} cluster, $\mathbf{z}_i=\mathbf{u}_3=(0,1)$, corresponding to $\mathbf{z}^{\star}_i=(0,0,1,0)$;
\item to both of them, $\mathbf{z}_i=\mathbf{u}_4=(1,1)$, the \emph{heir} cluster corresponding to $\mathbf{z}^{\star}_i=(0,0,0,1)$.
\end{itemize}

\end{document}